\documentclass[aps,prl,twocolumn,superscriptaddress]{revtex4}
\usepackage{amssymb,amsthm,amsmath,enumerate,url,graphicx,bbm}
\usepackage{amsbsy} % for bold greek letters
\usepackage[bold]{hhtensor}  
\usepackage{color}

\usepackage{subfigure}  % and subfigures

\usepackage{natbib}
\usepackage{algorithm}
\usepackage{algpseudocode}

\usepackage{multirow}

\theoremstyle{plain}

\theoremstyle{definition}

\newcommand{\KL}{D}

\newcommand{\Tr}{\mathrm{Tr}}
\newcommand{\cL}{\mathcal{L}}

\newcommand{\ket}[1]{\ensuremath{\left|#1\right\rangle}}
\newcommand{\bra}[1]{\ensuremath{\left\langle #1\right|}}
\newcommand{\braket}[2]{\ensuremath{\left\langle #1|#2\right\rangle}}
\newcommand{\ketbra}[2]{\ket{#1}\!\!\bra{#2}}
\newcommand{\expect}[1]{\ensuremath{\left\langle#1\right\rangle}}
\newcommand{\braopket}[3]{\ensuremath{\bra{#1}#2\ket{#3}}}
\newcommand{\proj}[1]{\ketbra{#1}{#1}}

\def\Id{1\!\mathrm{l}}

\newcommand{\rhohat}{\hat{\rho}}
\newcommand{\rhoML}{\rhohat_{\scriptscriptstyle\mathrm{ML}}}
\newcommand{\rhoLI}{\rhohat_{\scriptscriptstyle\mathrm{LI}}}

\newcommand{\dbar}{\overline{d}}
\newcommand{\dbarmax}{\dbar_{\mathrm{max}}}

\newcommand{\reals}{\mathbb{R}}

\usepackage{hyperref} % must be loaded last

\hypersetup{
  colorlinks   = true, %Colours links instead of ugly boxes
  urlcolor     = blue, %Colour for external hyperlinks
  linkcolor    = blue, %Colour of internal links
  citecolor   =  red %Colour of citations
}
\begin{document}

\title{Minimax quantum tomography:  the ultimate bounds on accuracy}
\author{Christopher Ferrie}
\affiliation{Center for Quantum Information and Control, University of New Mexico, Albuquerque, New Mexico, 87131-0001}
\affiliation{Centre for Engineered Quantum Systems, School of Physics, The University of Sydney, Sydney, NSW, Australia}

\author{Robin Blume-Kohout}
\affiliation{Sandia National Laboratories, Albuquerque, New Mexico, 87185}

\begin{abstract}
A \emph{minimax} estimator has the minimum possible error (``risk'') in the worst case.  We construct the first minimax estimators for quantum state tomography with relative entropy risk.  The minimax risk of non-adaptive tomography scales as $O(1/\sqrt{N})$, in contrast to that of classical probability estimation which is $O(1/N)$.  We trace this deficiency to \emph{sampling mismatch}:  future observations that determine risk may come from a different sample space than the past data that determine the estimate.  This makes minimax estimators very biased, and we propose a computationally tractable alternative with similar behavior in the worst case, but superior accuracy on most states.
\end{abstract}
\date{\today}

\maketitle

Quantum information processing relies on physical \emph{qubits} that store and process quantum information.  Testing and characterizing qubit devices is the business of quantum tomography \cite{paris2004quantum}, and \emph{quantum state tomography} in particular is used to estimate the quantum state (density matrix) $\rho$ produced by an initialization procedure.  Tomography comprises two steps:  (1) \emph{data gathering}, accomplished by measuring a ``quorum" of different observables on $N$ samples of $\rho$; and (2) an \emph{estimator} that maps the data to a final estimate $\rhohat$.  The goal, of course, is an accurate estimate -- we want $\rhohat$ to be ``close'' to the true state $\rho$, minimizing some error metric $d(\rho:\rhohat)$.  

One might thus expect that tomographers would choose an estimator that is optimal (or at least near-optimal) in accuracy.  Somewhat surprisingly, this is not done.  Although several estimators are known and used (linear inversion \cite{nielsen2010quantum}, maximum likelihood \cite{hradil1997quantum}, Bayesian mean \cite{blume2010optimal}, hedged maximum likelihood \cite{blume2010hedged}, $L_1$-regularization \cite{gross2010quantum}), none of them is known to have optimal pointwise accuracy \cite{foot:BME} for finite $N$.  In fact, we don't even know the ultimate bounds on accuracy, which makes it impossible to say which of these estimators (if any) are ``good enough''.

\begin{figure}[t!]
\centering
\begin{tabular}{l}
\textbf{(a)}  \vspace{-0.13in} \\ \includegraphics[width=1\columnwidth]{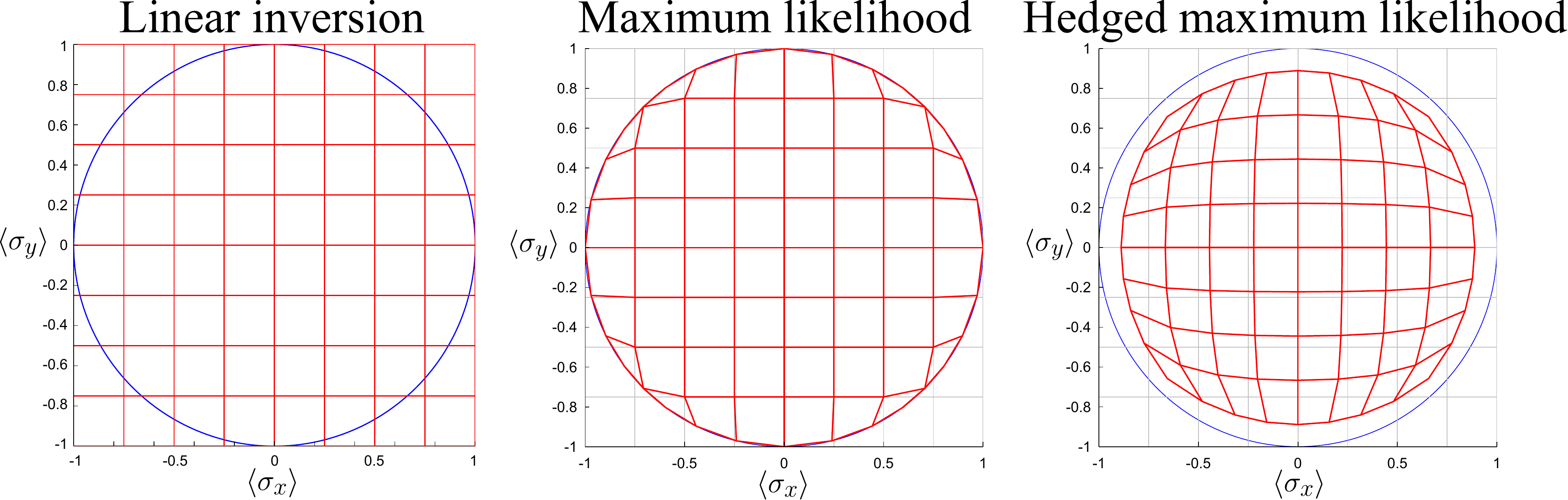} \\
\textbf{(b)}  \hspace{1in} Minimax Estimators
%\vspace{-0.1in} 
\\ \includegraphics[width=1\columnwidth]{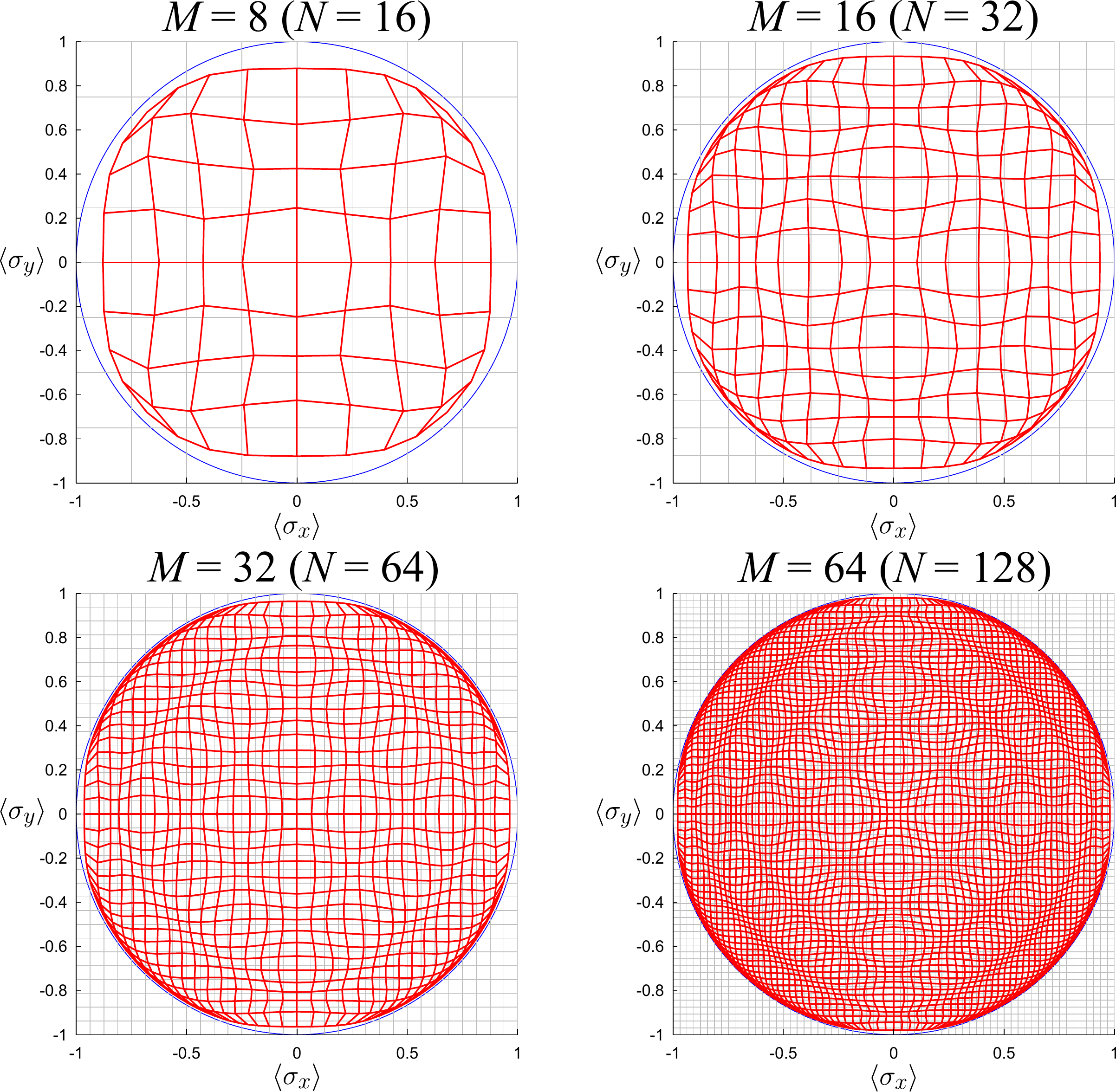}
\end{tabular}
\caption{\label{fig:minimax_estimators}
\textbf{Estimators for Pauli measurements on a rebit}, depicted as distortions of the ``linear inversion grid'' (see text after Eq. \ref{alpha_def}).  \textbf{(a)} Three standard estimators, each for $M=8$ measurements of $X$ and $Y$.  Each vertex of the red grid corresponds to an estimated density matrix.  Linear inversion estimates may lie outside ``Bloch disk'' of physical states.  MLE estimates are non-negative, while HML yields strictly positive estimates.
\textbf{(b)} Minimax estimators for $M= 8,16,32,64$ measurements of $X$ and $Y$ on a rebit.  They are locally biased, toward support points of the least favorable prior.}
\end{figure}

We remedy this embarrassing situation in the present Letter by constructing \emph{minimax} estimators (depicted in Fig. \ref{fig:minimax_estimators}; see detailed explanation after Eq. \ref{alpha_def}) with absolutely optimal performance.  These estimators are unwieldy, but (i) their performance yields tight upper bounds on accuracy, effectively delineating what ``good enough'' means, and (ii) their construction provides quite a lot of insight into the structure of the problem.  Armed with these results, we show that \emph{hedged maximum likelihood} (HML) is remarkably close to optimal, and outperforms minimax for most states (though of course its worst-case risk is higher).  We also identify a good value for the hedging parameter $\beta$ that appears in HML.

\noindent\textbf{Prerequisites:} Defining ``optimal'' requires making several choices.  For example, an optimal estimator for one error metric $d(\rho:\rhohat)$ is generally not optimal for a different metric $d'(\rho:\rhohat)$.  Here \cite{foot:other_loss_fcns}, we quantify inaccuracy by the \emph{quantum relative entropy},
\begin{equation}
d(\rho:\rhohat) = \KL(\rho||\rhohat) = \Tr\left[\rho(\log\rho-\log\rhohat)\right].
\end{equation}
Like its classical analogue, quantum relative entropy \cite{VedralRMP02} is a well-motivated measure of \emph{predictive} (and information-theoretic) inaccuracy \cite{blume2010optimal}.  It quantifies the expected cost, resulting from an imperfect estimate, of imperfectly predicting measurements of $\rho$'s diagonal basis (because this is the hardest measurement to predict accurately).

%Following common usage in the statistical and machine-learning literature \cite{lehmann1998theory}, 
An estimator's \emph{pointwise risk} is a function of the true state $\rho$ and is given by the average of $d(\rho:\rhohat)$ over all possible data sets $D$:
\begin{equation}\label{def:risk}
\dbar(\rho) = \sum_D{Pr(D|\rho)d(\rho:\rhohat(D))}.
\end{equation}
In the minimax paradigm, we quantify an estimator's accuracy by its \emph{worst-case} risk, $\dbarmax = \max_{\rho}\dbar(\rho)$.  The \emph{minimax risk} of the estimation problem is the minimum achievable risk (minimized over all possible estimators), and a \emph{minimax estimator} is one that achieves this bound.

In most inference problems, the sample space of possible observations (data) is fixed by the problem.  Not so in quantum tomography.  Quantum systems can be measured in many different and incomparable ways.  This is the single most significant difference between quantum and classical estimation.  This freedom is often removed in quantum problems by choosing the best or worst possible measurement (e.g., as in the definition of quantum relative entropy as the classical relative entropy of the most difficult-to-predict measurement).  This is usually not done in tomography, because the optimal measurements are far too difficult.  In this letter, we follow the majority of experiments and analyze tomography based on Pauli measurements on a single qubit.  However, we also prove analytic lower bounds on minimax risk that apply to \emph{any} non-adaptive measurement and any $d$-dimensional quantum system.  In some parts of our analysis, we use a \emph{rebit} -- a quantum system with a 2-dimensional \emph{real} Hilbert space, whose state space corresponds to the equatorial plane of the Bloch sphere -- as an easier-to-analyze proxy for a qubit.

\begin{figure}[t!]
\centering
\includegraphics[width=\columnwidth]{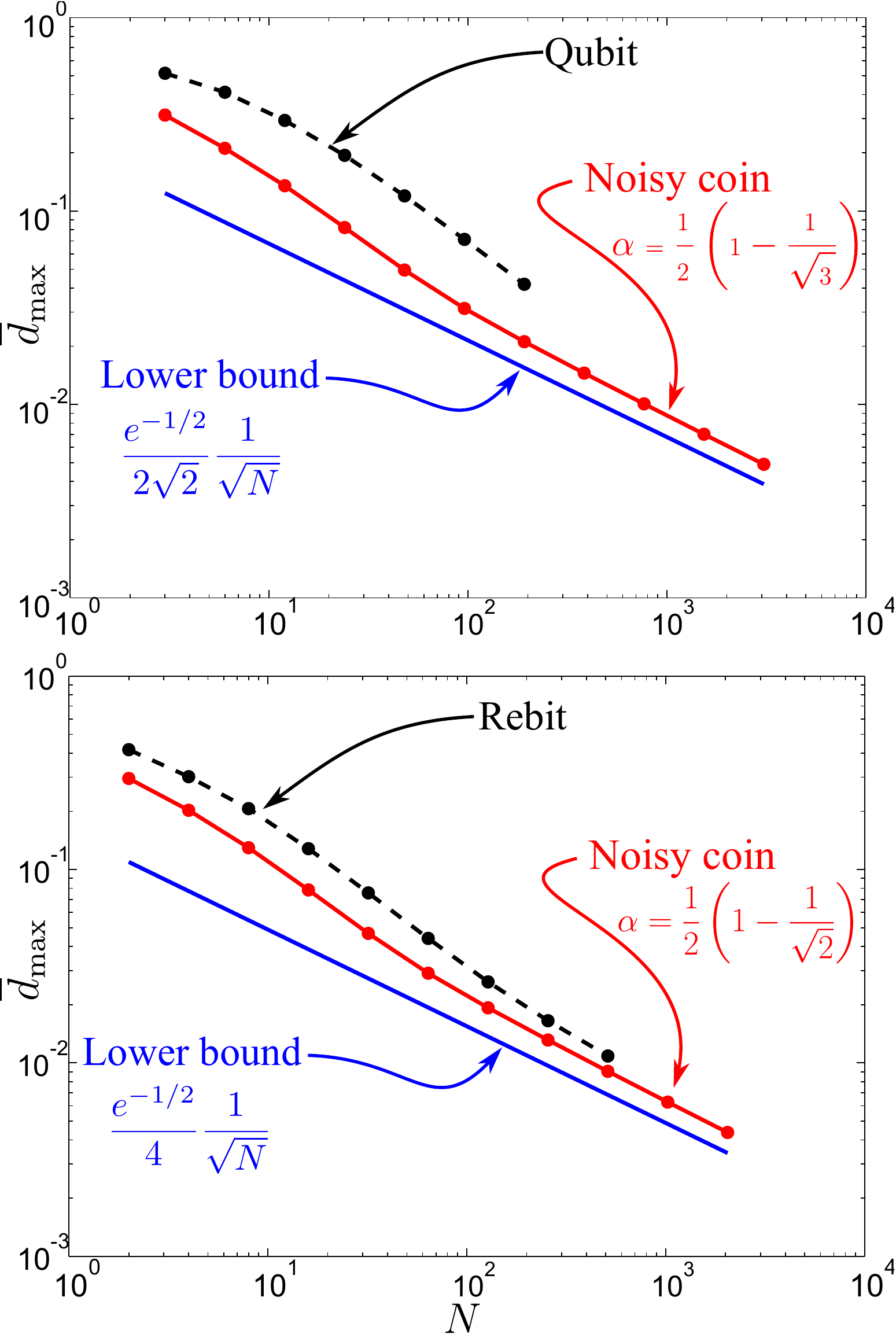}
\caption{\label{fig:NC_lower_bound}
\textbf{Numerical minimax risk for qubits, rebits, and noisy coins}.  Black curves show the risk of numerically constructed minimax estimators for \textbf{(a)} a qubit and \textbf{(b)} a rebit, as a function of the number of samples ($N$), up to the maximum that was numerically feasible.  Red curves illustrate the numerically-computed risk of ``noisy coin'' systems whose noise levels are chosen to match the effective ``noise'' of the qubit and the rebit (respectively).  Blue lines show the the lower bound given in Eq.~\eqref{noisy_coin_risk}.}
\end{figure}

\textbf{Minimax risk:}  The first main result of this Letter is a lower bound on the asymptotic ($N\to\infty$) minimax relative entropy risk of Pauli tomography on qubits and rebits,
\begin{equation}
\overline{d}_{\rm max} \geq  \frac{e^{-\frac12}}{4}\frac{\sqrt{D-1}}{\sqrt{N}},
\label{eq:MMRbound}
\end{equation}
where $D=2$ for rebits and $D=3$ for qubits.  Its $O(1/\sqrt{N})$ scaling contrasts sharply with the minimax risk of estimating a \emph{classical} bit, which is almost exactly $0.5/N$ \cite{rukhin1993minimax,krichevskiy1998laplace}.  We derive this bound below by mapping the minimax risk of qubit and rebit state tomography to a classical ``noisy coin'' model.  In Figure \ref{fig:NC_lower_bound}, we compare these bounds to numerical calculations of the minimax risk, for small $N$, of qubits, rebits, and noisy coins.

%We find that the minimax relative entropy risk for non-adaptive quantum tomography is asymptotically $O(1/\sqrt{N})$.  We prove this as a lower bound for any $d$-dimensional quantum system, and for any fixed choice of measurements (not just Pauli measurements).  For Pauli measurements on qubits, numerics for $N\leq192$ samples (Fig. \ref{fig:NC lower bound}a) indicate that the minimax risk is approximately $\dbarmax \approx 0.21/\sqrt{N}$.  For ``rebits" \cite{foot:rebit}, numerics on $N\leq512$ Pauli measurements (Fig. \ref{fig:NC lower bound}b) yield $\dbarmax \approx 0.15/\sqrt{N}$.  

%We obtained the numerical results shown in Fig. \ref{fig:NC lower bound} by numerical computation of explicit minimax estimators, which we will discuss shortly.  First, however, we wish to explain the $O(1/\sqrt{N})$ scaling with a simple analytical model.

A $d$-dimensional quantum state is analogous in many ways to a classical $d$-outcome probability distribution.  However, its minimax risk scales differently because of a phenomenon instrinsic to quantum tomography (though not uniquely quantum) that we call \emph{sampling mismatch}:  the sample space for the observed events is neither unique nor isomorphic to the underlying state space.  For example, the possible statistics for the three 2-outcome Pauli measurements on a qubit naturally define a cube, whereas the possible \emph{quantum} states form a sphere (the Bloch ball).  

Sampling mismatch can be reproduced in a simple classical model called the ``noisy coin'' \cite{Ferrie2012Estimating}.  It is a classical system with a 2-outcome sample space (i.e., a coin flip) where each observation is erroneous with known probability $\alpha$.  Sampling mismatch arises when we attempt to assign probabilities to future \emph{noiseless} observations using data from \emph{noisy} measurements.  The noisy coin's minimax risk is $O(1/\sqrt{N})$, because nearly-pure states are hard to estimate accurately from noisy statistics.  The corresponding minimax estimators are strongly biased toward nearly-pure states (see \cite{Ferrie2012Estimating} for details).  We are going to use a variant of the noisy coin model to bound the risk of tomography.

We define ``tomography'' thus: $N$ samples (copies) of a single-qubit state $\rho$ will be prepared; each sample will be measured independently (not jointly together with other samples) in a predefined fashion (not adaptively).  The $k$th sample is measured in an arbitrary basis, and this measurement can be described by a POVM (positive operator-valued measure) $\mathcal{M}_k = \{\Pi_k,\Id-\Pi_k\}$ whose outcomes have probabilities $\{q,1-q\}$ with $q=\Tr\Pi_k\rho$.  Based on the $N$ measurement results, we report a state $\rhohat$, and seek to minimize relative entropy cost.

Now, suppose that before analyzing the data (but after choosing the measurements!) we are told the eigenbasis of $\rho$.  This helps us (only $\rho$'s spectrum must be estimated), so the risk of spectrum estimation is a strict \emph{lower} bound on the risk of full tomography\cite{foot:errors}.  

We define $\{\ket{0},\ket{1}\}$ to be the eigenstates of $\rho$, and write
\begin{equation}
\rho = p\proj{0} + (1-p)\proj{1}.
\end{equation}
Now, we need only estimate $p\in[0,1]$.  This parameter manifold is identical to that of a coin.  Furthermore, the quantum relative entropy between two diagonal density matrices is identical to the classical relative entropy between the corresponding distributions.  So, since $\rho$'s eigenbasis is known, estimating $\rho$ is identical to estimating the bias of a coin.  However, unless the eigenbases of $\rho$ and the $\Pi_k$ happen to coincide, the measurement data obtained from the $N$ samples of $\rho$ are not ``noiseless''.  Even if $p=0$ (i.e., $\rho$ is pure), the data remain somewhat random.  The probability of observing $\Pi_k$ is not $p$, but
\begin{eqnarray*}
q &=& p\braopket{0}{\Pi_k}{0} + (1-p)\braopket{1}{\Pi_k}{1} \\
&=& p(1-2\alpha_k) + \alpha_k
\end{eqnarray*}
where the \emph{effective noise} in sample $k$ is
\begin{equation} \label{def_alphak}
\alpha_k = \braopket{1}{\Pi_k}{1}^2.
\end{equation}
We can model this situation perfectly by a noisy coin (as in Ref. \cite{Ferrie2012Estimating}) where each observation fails with a different error probability.  The error probability for the $k$th sample is $\alpha_k$.  In the appendix, we bound this estimation problem's minimax risk by
\begin{equation}\label{noisy_coin_risk}
\dbar_{\rm max}\geq \frac{e^{-\frac12}}{2\sqrt{\bar\beta}}\frac1{\sqrt N},
\end{equation}
where $\bar\beta$ is the average \emph{resolution} provided by the $N$ noisy samples:
\begin{equation}\label{alpha_def}
\bar \beta = \frac{1}{N}\sum_{k=1}^N{\beta_k} = \frac{1}{N}\sum_{k=1}^N\frac{(1-2\alpha_k)^2}{\alpha_k(1-\alpha_k)}.
\end{equation}

For any fixed measurement strategy -- e.g., the standard one where $N/3$ samples are measured in the $X,Y,Z$ bases -- the \emph{maximum} risk occurs when we choose the eigenbasis of $\rho$ to maximize $\bar\beta$ in Eq. \ref{alpha_def}.  This ``least favorable'' basis is the one that lies as far as possible from all measured bases.  For a rebit, it lies halfway between the $X$ and $Z$ bases, and $\alpha_k=\frac12(1-1/\sqrt{2})$.  For a qubit, it is the geometric mean of the $X$, $Y$, and $Z$ bases, and $\alpha_k = \frac12(1-1/\sqrt{3})$.  Inserting these values for $\alpha_k$ yields the final bound given in Eq. \ref{eq:MMRbound}.

This argument applies (qualitatively) to tomography on any finite-dimensional system with any discrete POVM. As long as no samples are measured in a basis that diagonalizes $\rho$, the minimax risk scales as $O(1/\sqrt{N})$ (although the prefactor will vary).  However, if any nonvanishing fraction of the $N$ samples are measured in a basis that diagonalizes $\rho$, then Eq. \ref{noisy_coin_risk} no longer applies.  Thus, continuous POVMs such as the unitarily invariant Haar-uniform rank-1 POVM (a.k.a. the uniform POVM), require a slightly different argument.  In the appendix, we prove that even in this case, the minimax risk is lower bounded by $O\left((N \log N)^{-1/2}\right)$.

\textbf{Estimators:}  
To confirm the bound given by Eq. \ref{eq:MMRbound} and explore minimax risk at small $N$, we use numerics to find minimax estimators.  An estimator is a map from the set of all possible datasets into the set of density matrices.  The outcomes of the measurement(s) performed are represented by a set of positive operators $\{E_k\}$, and the data themselves by a set of frequencies $D = \{n_k\}$.  For qubit Pauli tomography, the data comprise $M=N/3$ samples each of $\sigma_x$, $\sigma_y$, and $\sigma_z$ measurements; for rebits, they comprise $M=N/2$ samples each of $\sigma_x$ and $\sigma_y$ measurements. 

We used numerical optimization (over the set of possible estimators) to find minimax estimators.  The algorithms are described in the appendix.  In Figure \ref{fig:minimax_estimators}, we depict the resulting estimators, and compare them to three canonical estimators:
\begin{enumerate}
\item \textbf{Linear inversion} ($\rhoLI$):  The first tomographic estimator, it is obtained by equating each probability $\Pr(k|\rhoLI) = \Tr E_k\rhoLI$ to its observed frequency $\frac{n_k}{M}$.
\item \textbf{Maximum likelihood} ($\rhoML$):  MLE assigns the density matrix that maximizes the probability of the observed data (the likelihood), $\cL(\rho) =\ Pr(D|\rho) = \prod_k\left[\Tr(E_k\rho)^{n_k}\right]$.  

\item \textbf{Hedged maximum likelihood} ($\hat \rho_{{\rm HML},\beta}$):  The HML estimator maximizes the product of $\cL(\rho)$ and a ``hedging function'' $h(\rho) = \det(\rho)^\beta$.  This function is strictly convex and vanishes for rank-deficient states, so the HML estimate is always full-rank.
%\item \textbf{Minimax} ($\hat\rho_{\text{opt}}$): the estimator we discussed above which solves the optimization problem:
%\begin{equation}\label{eq:minimax def}
%\hat\rho_{\text{opt}} = \argmin_{\hat\rho}\max_\rho\sum_D{Pr(D|\rho)d(\rho:\rhohat(D))}.
%\end{equation}
\end{enumerate}
To simplify visualization, we depict \emph{rebit} estimators, which are qualitatively similar to qubit estimators and easier to depict.  A rebit estimator is a map from datasets to Bloch vectors, as $\hat\rho:\{0,\ldots,M\}^2\to\reals^2$.  We use the linear inversion estimator as a reference.  As a linear map from the 2-dimensional space of datasets ($\{0\ldots M\}^2$) and the 2-dimensional space of rebit states (the unit disc in $\reals^2$), the linear inversion estimator is represented by a uniform grid on the ``Bloch square'' (Fig. \ref{fig:minimax_estimators}a).  Every \emph{other} estimator is represented as a distortion of this grid.  The vertices of the grid are estimates $\rhohat$, and the position of such a vertex within the grid indicates what dataset it came from.  

Minimax estimators for $N=16,32,64$ and $128$ (total) Pauli measurements on a rebit are shown in Figure \ref{fig:minimax_estimators}b.  The most striking feature of these estimators is a pronounced ``ripple'' phenomenon.  This is not a numerical artifact.  Instead, it represents a consistent bias toward certain discrete points within the state space (support points of the least favorable prior -- see Fig. \ref{fig:2LFPs} in the appendix), which can be identified in Figure \ref{fig:minimax_estimators} as regions where the grid lines cluster together.  The minimax estimator demonstrates this bias because these points are, in a particular sense, the most difficult to estimate accurately.

%\begin{figure}[ht]
%\centering
%\includegraphics[width=1\columnwidth]{rebitN8estimators}
%\caption{\label{fig:linear_grid_N8}
%On the left, the ``linear inversion grid'' for $M=8$ Pauli measurements on a rebit.  Each vertex corresponds to an estimated density matrix.  Note that, for some data sets, the estimated density matrix lies outside the allowed states in the ``Bloch circle''.  In the middle, a depiction of how the maximum likelihood estimator deforms the linear inversion grid.  Note that all states which linear inversion estimates to be outside the state space have been truncated down to the boundary.  Finally, on the right, we show the HML estimator with $\beta=1/2$.  The hedging has the effect of pushing all estimates toward the center of the Bloch circle.  This effect increases with $\beta$.}
%\end{figure}

\begin{figure}[t!]
\centering
\includegraphics[width=\columnwidth]{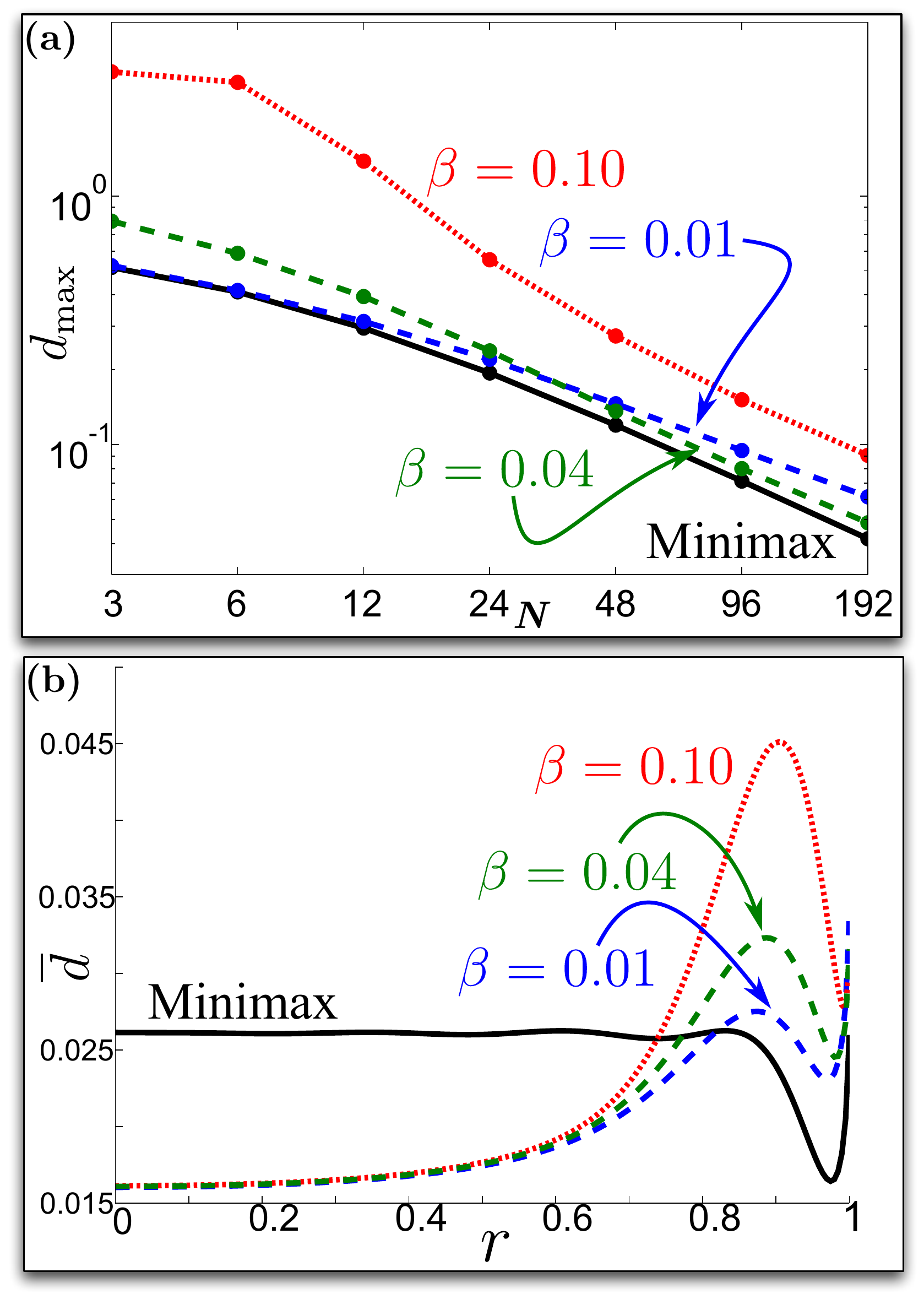}
\caption{\label{fig:PointwiseRisk}
\textbf{Maximum and pointwise risk of minimax and HML estimators}.  Plot \textbf{(a)} shows the maximum risk, for \emph{qubit} tomography, of the minimax estimator and three different HML estimators ($\beta=0.01, 0.04,0.10$) for $N\leq192$ samples distributed equally among the 3 Pauli bases.  Plot \textbf{(b)} shows the pointwise risk, along the axis oriented at 45 degrees to both $X$ and $Y$, of the same estimators for $N=128$ samples for a \emph{rebit} (this minimax estimator is depicted in Fig. \ref{fig:minimax_estimators}b).  The two local maxima of $\dbar(\rho)$ are at $r=1$ and $r\approx1-\frac{1}{\sqrt{N}}$.  Choosing $\beta\approx0.04$ balances these risks, and is therefore minimax among HML estimators.  This optimal HML estimator comes very close to matching the worst-case performance of the minimax estimator, and outperforms it dramatically in the interior of the state space.}
\end{figure}

\textbf{Improving on Minimax}:  The minimax criterion is an elegant concept, but a dangerous one.  In its single-minded quest to improve the \emph{maximum} risk, it has no concern for the pointwise risk at states that are ``easier'' to estimate.  In such regions, it may incur extreme bias and inaccuracy, for the sole purpose of achieving a tiny reduction in the maximum risk.  For quantum tomography, this effect become extreme.  While $O(1/N)$ risk can be achieved on all full-rank states, the risk is unavoidably $O(1/\sqrt{N})$ near the boundary.  Our numerical experiments confirm that the minimax estimator's pointwise risk is $O(1/\sqrt{N})$ everywhere, whereas other estimators easily achieve $O(1/N)$ risk in the interior of the Bloch sphere (Fig. \ref{fig:PointwiseRisk}b).  If $\rho$ really was selected adversarially, then minimax would be a wise strategy.  But in realistic cases, we would prefer an estimator that achieved $O(1/N)$ scaling where possible, even at the cost of slightly worse \emph{worst-case} behavior.

A good estimator should achieve $O(1/N)$ risk in the interior, while coming as close as possible to minimax performance near the boundary.  The maximum likelihood estimator (MLE) is disqualified because its pointwise expected risk is uniformly infinite (it has nonzero probability of returning a rank-deficient estimate for every $\rho$, so $\dbar(\rho)=\infty$).  However, \emph{hedged maximum likelihood} (HML) does not have this behavior.  Introduced in Ref. \cite{blume2010hedged} as a full-rank alternative to MLE, HML generalizes classical ``add-$\beta$'' estimators.  Like them, it never assigns zero probabilities, and has a adjustable parameter $\beta$ that governs how much it avoids zero eigenvalues.  Classical ``add-$\beta$'' estimators are very nearly minimax (for $\beta\approx1/2$), which suggests that HML estimators might have similar near-optimality properties.

All HML estimators have good behavior ($O(1/N)$ pointwise risk) in the interior, so we are free to define the ``optimal'' $\beta$ by minimax (among HML estimators).  As illustrated in Fig. \ref{fig:PointwiseRisk}b, an HML estimator's pointwise risk has local maxima at the boundary (pure states) and/or at a slightly depolarized state (with purity $\sim1-1/\sqrt{N}$).  To minimize its maximum, we choose $\beta$ to equalize the risk at these two local maxima.  The asymptotically optimal $\beta$ for the noisy coin model was shown in Ref. \cite{Ferrie2012Estimating} to be $\beta_{\mathrm{optimal}} \approx 0.0389$, and our numerics confirm that $\beta\approx0.04$ is optimal to within the available numerical precision for rebit tomography as well (Fig. \ref{fig:PointwiseRisk}b; qubit results for smaller $N$ are not shown, but confirm that $\beta\approx0.04$ has nearly-minimax performance).

For this near-optimal value of $\beta$, HML compares favorably with minimax estimators.  Its worst-case risk is very close to the minimax risk (Fig. \ref{fig:PointwiseRisk}a), and it dramatically outperforms minimax in the interior of the state space (Fig. \ref{fig:PointwiseRisk}b).  So while optimal hedging estimators do not offer strictly optimal performance by any criterion, they are (i) easy to specify and calculate, (ii) close to minimax, and (ii) more accurate than minimax estimators for almost all states $\rho$.  We do not why the optimal $\beta$ is so different for noiseless coins ($\approx 0.5$) and for qubits/rebits/noisy coins ($\approx0.04$), but it suggests fundamental differences between noiselessly sampled systems and those (like qubits and noisy coins) where sampling mismatch is important.

\acknowledgments{CF was supported by National Science Foundation grant number PHY-1212445, the Canadian Government through the NSERC PDF program, the IARPA MQCO program, the ARC via EQuS project number CE11001013, and by the US Army Research Office grant numbers W911NF-14-1-0098 and W911NF-14-1-0103. Sandia National Laboratories is a multi-program laboratory operated by Sandia Corporation, a wholly owned subsidiary of Lockheed Martin Corporation, for the U.S. Department of Energy's National Nuclear Security Administration under contract DE-AC04-94AL85000.}
%=================================================
% References
% NOTE: if someone is reading this and knows how to have BibTeX not make this so messy, please tell me!
%=================================================

%==================================================
% Appendix
%==================================================

\onecolumngrid
\appendix

\section{The minimax risk of a noisy coin}

In this appendix we show that the minimax risk of estimating the bias of a \emph{noisy} coin is $O(1/\sqrt N)$ (in contrast to the $O(1/N)$ minimax risk for a noiselessly observed coin), and derive a simple lower bound on it.

Now, suppose a coin with bias $p=\mathrm{Pr}(\mathrm{``heads''})$ is flipped $N$ times and a sequence of binary outcomes $\vec{n} = \{n_k\}$ are recorded.  But these observations are unreliable; each outcome is recorded incorrectly with trial-dependent probabilities $\vec{\alpha} = \{\alpha_k\}$ (all taken from the interval $\left[0,\frac12\right)$).
The distribution of the outcomes is
\begin{equation}\label{likelihood function}
\Pr(\vec{n}|p,\vec{\alpha})=\prod_{k=1}^N \Pr(n_k|p,\alpha_k) = \prod_{k=1}^N q_k^{n_k}(1-q_k)^{1-n_k},
\end{equation}
where the probability of observing ``heads'' on trial $k$ is not $p$, but 
\begin{equation}
q_k(p) = \alpha_k + p(1-2\alpha_k) = p + \alpha_k\left(1-2p\right).
\end{equation}
We recover a standard noiseless coin when $\alpha_k=0$ for all $k$.

For each prior distribution $\mu(p)$, the estimator with the smallest risk (expected cost) is the \emph{Bayes estimator} for $\mu(p)$, and its risk is the \emph{Bayes risk} of $\mu(p)$.  Bayes estimators need not be simple, but because relative entropy is a \emph{Bregman divergence}, the Bayes estimator is always the mean of the posterior distribution (obtained via Bayes' Rule) \cite{blume2010optimal}. The prior with the highest Bayes risk is the \emph{least favorable} prior, and its risk is the minimax risk.  Thus, the Bayes risk of any prior is a lower bound for the minimax risk, which suggests an obvious variational approach to bounding the minimax risk by choosing a prior whose risk is high but easy to calculate.  Obviously, some priors have very low risk [e.g., $\mu(p) = \delta(p-p_0)$], and provide useless lower bounds.  A common approach is to use the uniform (Lebesgue) prior, but for the noisy coin this prior actually has rather low [$O(1/N)$] risk.  So instead, we consider the set of \emph{bimodal priors},
\begin{equation}
\pi(p) = \frac{\delta(p-p_0) + \delta(p-p_1)}{2}.
\end{equation}
(Varying the weights yields a slightly less favorable prior, but doesn't change the asymptotic scaling).  We will choose $p_0=0$ and $p_1 \approx 1/\sqrt{N}$.

The risk of $\pi$ is given by $\overline{d}(\pi) = [\overline{d}(0) + \overline{d}(p_1)]/2$, and by observing that $\overline{d}(p_1)\geq0$ we obtain the lower bound
\begin{equation}\label{nc_bimodal}
\overline{d}(\pi)\geq \frac 12\overline{d}(0)= \frac12 \mathbb E_{\vec{n}|p=0}[D(0\|\hat p(\vec{n}))],
\end{equation}
where the Bayes estimator is the posterior mean, given by
\begin{equation}
\hat p (\vec{n}) = \frac{p_1\Pr(\vec{n}|p_1)}{\Pr(\vec{n}|p_1)+\Pr(\vec{n}|0)} = \frac{p_1}{1+\Lambda(\vec{n})},
\end{equation}
in terms of the \emph{likelihood ratio}
\begin{equation}
\Lambda(\vec{n}) = \frac{\Pr(\vec{n}|0)}{\Pr(\vec{n}|p_1)}.
\end{equation}
We can lower-bound the relative entropy term by $D(0\|\hat p)=-\log(1-\hat{p}) \geq \hat{p}$, so
\begin{equation}
\overline{d}(0)\geq \mathbb E_{\vec{n}|p=0}[\hat p(\vec{n})] = p_1\mathbb E_{\vec{n}|p=0}\left[\frac{1}{1+\Lambda(\vec n)}\right].
\end{equation}
If we define $\lambda(\vec{n}) =  -2\log\Lambda(\vec n)$ and apply Jensen's inequality, we obtain 
\begin{equation}
\overline{d}(0) \geq p_1 e^{-\frac12\mathbb E_{\vec{n}|p=0}\left[\lambda(\vec n)\right]}.
\end{equation}
Next, we perform a Taylor expansion of the expectation $E_{\vec{n}|p=0}\left[\lambda(\vec n)\right]$ around $p_1=0$. The derivatives of the likelihood function \eqref{likelihood function} are
\begin{align}
\frac{\partial}{\partial p_1} \log \Pr(\vec{n}|p_1) & = \sum_k \left(\frac{n_k(1-2\alpha_k)}{q_k}-\frac{(1-n_k)(1-2\alpha_k)}{1-q_k}\right),\\
\frac{\partial^2}{\partial p_1^2} \log \Pr(\vec{n}|p_1) & = \sum_k \left(-\frac{n_k(1-2\alpha_k)^2}{q_k^2}-\frac{(1-n_k)(1-2\alpha_k)^2}{(1-q_k)^2}\right).
\end{align}
Evaluating these at $p_1=0$ and taking the expectation $\mathbb E_{\vec{n}|p=o}[n_k] = \alpha_k$, we have
\begin{align}
\mathbb E_{\vec{n}|p=o}\left[\frac{\partial}{\partial p_1} \log \Pr(\vec{n}|p_1)\right]_{p_1=0} & = 0,\\
\mathbb E_{\vec{n}|p=o}\left[\frac{\partial^2}{\partial p_1^2} \log \Pr(\vec{n}|p_1)\right]_{p_1=0} & = \sum_k \frac{(1-2\alpha_k)^2}{\alpha_k(1-\alpha_k)}.
\end{align}
Putting everything together in the Taylor series, we obtain 
\begin{equation}\label{Eseries}
\mathbb E_{\vec{n}|p=0}\left[\lambda(\vec n)\right] = \sum_{k=1}^N\frac{(1-2\alpha_k)^2}{\alpha_k(1-\alpha_k)}p_1^2 + O(p_1^3),
\end{equation}
where the $O(p_1^3)$ term does not scale with $N$ w/r.t. the leading order term.

To simplify this quantity, we define the per-sample ``resolution'' $\beta_k$,
$$
\beta_k = \frac{(1-2\alpha_k)^2}{\alpha_k(1-\alpha_k)},
$$
(which, at least for small $\alpha_k$, is approximately $1/\alpha_k$).  The expectation value in Eq. \ref{Eseries} can be written concisely in terms of the average $\beta$, $\overline{\beta} = N^{-1}\sum_{k=1}^N{\beta_k}$, as
$$
\mathbb E_{\vec{n}|p=0}\left[\lambda(\vec n)\right]\sim N \overline{\beta} p_1^2.
$$
Finally, we set
\begin{equation}\label{pdef}
p_1 = \frac{1}{\sqrt{\overline{\beta}}}\frac{1}{\sqrt N},
\end{equation}
which ensures that $p_1\to 0$ as $N\to\infty$ and justifies truncating the series expansion Eq. \ref{Eseries} above at leading order.  This yields a lower bound on the minimax risk of
\begin{equation}
\overline{d}_{\rm max} \geq \overline{d}(\pi) \geq \frac{1}{2\sqrt{e\overline{\beta}}}\frac1{\sqrt N}.
\end{equation}
It is worth noting that the risk is \emph{not} determined by the average value of $\alpha$ (the per-sample noise probability), but by the average of $\beta$, which behaves roughly like $1/\alpha$.  In particular, if any constant fraction of the samples are observed noiselessly, then those samples have $\beta=\infty$, and they dominate the minimax risk -- $\overline{\beta}\to\infty$, and the minimax risk collapses to $O(1/N)$, as is appropriate for a noiseless coin.

\section{Minimax risk for quantum tomography}

In this section, we derive a lower bound for the minimax risk of qubit state estimation using the same framework that we used for the noisy coin.  The difficulty in doing this is that a qubit's state space (the Bloch sphere) is more complex than that of a coin -- instead of a single parameter ($p$) there are three ($x,y,z$).  However, the minimax risk is dominated by (i) states $\rho$ that are very close to pure, and (ii) errors in estimating the \emph{spectrum} of $\rho$ (rather than errors in its eigenvectors, which contribute much less to the risk).  This observation allows us to simplify the analysis greatly by choosing a bimodal prior for the qubit, supported on two states that differ only in their eigenvalues.  In this circumstance, each measurement provides information equivalent (in its effect on the final estimate) to a noisy coin flip whose noisyness depends on what was measured (or, most generally, on which outcome was observed).  Because we have chosen a very simple prior that is not least favorable, our analysis only guarantees a lower bound.  However, it captures the dominant component of the minimax risk, and (for such a simple model) turns out to be surprisingly close to tight.

Suppose we are given $N$ samples of a qubit state $\rho$.  The state is drawn from a bimodal prior supported on (i) a pure state $\rho_0 = \proj{\psi}$, and (ii) a slightly more mixed state $\rho_1 = (1-p_1)\proj{\psi} + p_1(\Id-\proj{\psi})$:
\begin{equation}\label{qubit_bimodal}
\pi(\rho) = \frac12\left(\delta(\rho-\rho_0) + \delta(\rho-\rho_1)\right).
\end{equation}
We will specify $\ket\psi$ and $p_1\approx1/\sqrt{N}$ later.  Each sample is measured in some basis; on the $k$th sample we perform the [orthogonal basis] POVM $\{\ketbra{\phi_k}{\phi_k},\mathbbm 1-\ketbra{\phi_k}{\phi_k}\}$, and list the outcomes as a binary vector $\vec n:=\{n_k\}$.

The likelihood function for a single observation is\begin{align}
\Pr(n_k|\rho_0) &= |\braket{\phi_k}{\psi}|^2=:\alpha_k\\
\Pr(n_k|\rho_1) &= (1-2p_1)|\braket{\phi_k}{\psi}|^2+p_1=(1-2\alpha_k)p_1 +\alpha_k.
\end{align}
These are identical to the likelihoods for the noisy coin.  The Bayes estimator is
\begin{equation}\label{qubit_bimodal_bayes}
\hat \rho (\vec{n}) =\frac{[\Pr(\vec{n}|\rho_0)+(1-2p_1)\Pr(\vec{n}|\rho_1)]\ketbra\psi\psi +p_1)\Pr(\vec{n}|\rho_1)\Id}{\Pr(\vec{n}|\rho_0)+\Pr(\vec{n}|\rho_1)}.
\end{equation}
Now, to compute the expected risk, we observe that the Bayes estimate is always of the form $\hat\rho = \alpha \ketbra\psi\psi + \beta \mathbbm 1$, with 
\begin{equation}
\alpha = \frac{[\Pr(\vec{n}|\rho_0)+(1-2p_1)\Pr(\vec{n}|\rho_1)]}{\Pr(\vec{n}|\rho_0)+\Pr(\vec{n}|\rho_1)},\ \beta = \frac{p_1\Pr(\vec{n}|\rho_1)}{\Pr(\vec{n}|\rho_0)+\Pr(\vec{n}|\rho_1)}.
\end{equation}
and for any such mixture $\sigma = \alpha \ketbra\psi\psi + \beta \mathbbm 1$, the relative entropy can be computed as
\begin{align}
D\left(\ketbra\psi\psi\|\sigma\right) &= - \braopket{\psi}{\log\sigma}{\psi} \\
&= -\braopket{\psi}{\log(\alpha+\beta)\ketbra\psi\psi + \log\beta (\mathbbm 1 - \ketbra\psi\psi)}{\psi}\\
&= -\log(\alpha+\beta).
\end{align}
Thus, in the limit of $p_1\to0$ and $N\to\infty$, the risk \emph{given} that $\rho=\rho_0$ is given by
\begin{align}
D(\rho_0|\hat\rho(\vec{n})) &= -\log\left[\frac{\Pr(\vec{n}|\rho_0)+\Pr(\vec{n}|\rho_1)\left(1-p_1\right)}{\Pr(\vec{n}|\rho_0)+\Pr(\vec{n}|\rho_1)}\right]\\
&= -\log\left[1-\frac{p_1\Pr(\vec{n}|\rho_1)}{\Pr(\vec{n}|\rho_0)+\Pr(\vec{n}|\rho_1)}\right]\\
&= p_1 \frac{\Pr(\vec{n}|\rho_1)}{\Pr(\vec{n}|\rho_0)+\Pr(\vec{n}|\rho_1)} + O(p_1^2).
\end{align}
This is identical to the risk of the noisy coin.

As in Eq. \ref{pdef}, we choose
\begin{equation}
p_1 = \frac{1}{\sqrt{\bar\beta}}\frac{1}{\sqrt N},
\end{equation}
where $\bar\beta$ is defined in Equation \ref{alpha_def} as
\begin{equation*}
\bar \beta = \frac{1}{N}\sum_{k=1}^N{\beta_k} = \frac{1}{N}\sum_{k=1}^N\frac{(1-2\alpha_k)^2}{\alpha_k(1-\alpha_k)}.
\end{equation*}
This yields a near-final lower bound of
\begin{equation}
\overline{d}_{\rm max} \geq r(\pi) \geq \frac{e^{-\frac12}}{2\sqrt{\bar\beta}}\frac1{\sqrt N}.
\end{equation}
To obtain a concrete lower bound on the risk, we must choose $\proj{\psi}$.  To ensure that the average resolution $\bar\beta$ is as small as possible near $\ket\psi$, we want $\alpha_k$ to be uniformly large.  For the case of a qubit or a rebit, the solution is to pick the state ``furthest away'' from all the measurement axes, which yields $\alpha_k = \frac12\left(1-\frac1{\sqrt D}\right)$ where $D=2$ for a rebit and $D=3$ for a qubit.  This yields the simple result $\bar{\beta} = 4/(D-1)$, and therefore
\begin{equation}
\overline{d}_{\rm max} \geq r(\pi) \geq \frac{e^{-\frac12}}{4}\frac{\sqrt{D-1}}{\sqrt{N}}.
\end{equation}
This argument can be extended to any discrete POVM, by choosing $\ket\psi$ so that it is not orthogonal to any effect of the POVM.  Then $\alpha_k$ for each $k$ will be lower-bounded by the minimum overlap of $\ket\psi$ with any effect, and $\bar\beta$ will be finite, and so the minimax risk will scale as $1/\sqrt{N}$.

But, by exactly the same argument, the \emph{best} nonadaptive tomographic measurement must be unitarily symmetric.  And since it must also be rank-1, it is the Haar-uniform POVM whose effects include all pure states $\proj\phi$ with the unitarily invariant measure.  The analysis given so far breaks down for this uniform POVM, because no matter what $\ket\psi$ we choose, the measurement has effects that diagonalize it.  The effective ``noise''
\begin{equation}
\Pr(n_k|\rho_0) = |\braket{\phi_k}{\psi}|^2=:\alpha_k
\end{equation}
is distributed uniformly over $[0,1]$.  If we attempt to replace the sum
\begin{equation}
\sum_{k=1}^{N} \frac{(1-2\alpha_k)^2}{\alpha_k(1-\alpha_k)}= : N \overline{\beta},
\end{equation}
with its average, then the integral diverges and our lower bound collapses to $\overline{d}\geq0$.

Instead, we observe that the Haar-uniform POVM can be described as a two-step process: (1) choose a Haar uniform-basis, and (2) measure in that basis.  So, a tomography experiment involving $N$ samples can be described by a sequence $[\alpha_1,\alpha_2,\ldots \alpha_N]$, in which each $\alpha_k$ is drawn from the uniform distribution over $[0,1]$.  The minimax risk is the [probability-weighted] average over all such sequences.  We divide them into two subsets:  those in which all the $\{\alpha_k\}$ lie in the interval
$$\alpha_k \in \left[\frac{1}{2N},1-\frac{1}{2N}\right]$$
and those in which at least one does not.

The probability that any given $\alpha_k$ lies outside the interval is exactly $1/N$, so \emph{all} of them lie within it with probability
$$p = \left(1-\frac{1}{N}\right)^N \geq \frac{1}{e}.$$
Conditional on all the $\{\alpha_k\}$ lying within the interval, the minimax risk can be lower bounded by integrating $\beta$ over the interval, which yields
\begin{equation}
\overline{\beta} = 2\log N + O(1).
\end{equation}
This happens with probability at least $1/e$, so a lower bound on the minimax risk for the Haar-uniform POVM (as $N\to\infty$) is
\begin{equation}
\overline{d} \geq \frac{1}{(2e)^{3/2}}\frac{1}{\sqrt{N\log N}}.
\end{equation}

\section{Least favorable priors}

The ``Optimization Toolbox'' in MATLAB 2011a, for example, contains a method \texttt{fminimax} which directly solves the optimization problem we are interested in.
%---Eq.~\eqref{eq:minimax def}.
However, finding minimax estimators by brute force seems impossibly difficult.  There are uncountably many estimators; each one is a density-matrix valued function on the set of all possible datasets.  Each estimator's performance is quantified by maximizing its risk profile $\dbar(\rho)$ over all density matrices $\rho$.  Even computing the maximum risk of a single specified estimator is nontrivial; finding its minimum over the uncountable set of \emph{all} estimators seems intractable.  

Fortunately, we have some useful mathematical tools that simplify matters greatly (see, for example, \cite{lehmann1998theory}):
\begin{enumerate}
\item The minimax estimator is \emph{also} the Bayes estimator for some measure.  This fact is called \emph{Minimax-Bayes duality}, and the measure in question is called a \emph{least favorable prior} (LFP).
\item Relative entropy is a Bregman divergence (a.k.a. strictly proper scoring rule), and therefore the Bayes estimator for any given measure $\mu$ is Bayesian mean estimation (BME).
\item The least favorable priors for this problem are (empirically) always \emph{discrete}, with a finite number of support points.  This is not proven, but it is often the case in similar problems, and is easy to verify numerically for this problem.
\end{enumerate}
Minimax-Bayes duality is enormously helpful, both as a technical tool and as an aid to problem-solving.  The reasoning behind this duality is fairly straightforward:

\begin{enumerate}
\item Any estimator involves trade-offs in accuracy, which are quantified by its risk profile $\dbar(\rho)$.  For example, the constant estimator $\rhohat(D) = \rho_0$ is exceptionally accurate \emph{if} the true state happens to be $\rho_0$!  That is, $\dbar(\rho_0) = 0$.  No other estimator can match its accuracy at $\rho_0$.  But there is a price to be paid; $\dbar(\rho)$ is dreadfully high for any state $\rho$ that is far from $\rho_0$.
\item Averaging $\dbar(\rho)$ over a measure $\mu$ quantifies these tradeoffs.  In order to minimize that average, the Bayes estimator for $\mu$ must achieve fairly low expected risk in regions where $\mu$ is concentrated, but can tolerate high risk where $\mu$ is sparse.  
\item If we consider the Bayes estimator for a specific measure $\mu_0$, its risk profile $\dbar(\rho)$ will typically be non-constant -- so it will have at least one maximum, which we denote $\rho_0$.  Now suppose that we modify $\mu_0$ (to $\mu'$) by slightly increasing the probability density around $\rho_0$.  The new measure $\mu'$ will have a higher Bayes risk (since $\rho_0$ has higher-than-average risk, and is now slightly more probable).  But the Bayes estimator for $\mu'$ will be slightly different as well; it will achieve a \emph{lower} value of $\dbar(\rho_0)$ because by increasing the probability of $\rho_0$ we have increased the value of achieving low risk at $\rho_0$.
\item Iterating this process defines a flow -- probability flows towards high-risk states (decreasing their expected risk) and away from low-risk states (increasing their expected risk).  Every step in this flow defines a new prior (and its associated Bayes estimator) with higher Bayes risk and lower maximum risk.
\item If $\mu$ is a stationary point of this flow, then the risk profile of its Bayes estimator $\rhohat_\mu(D)$ is: (i) equal to a constant $C$ on the support of $\mu$, and (ii) no greater than $C$ at every point \emph{not} in the support of $\mu$.  This estimator is necessarily minimax, because:
\begin{itemize}
\item No estimator can achieve lower \emph{average} risk on $\mu$ (by the definition of Bayes estimator),
\item So no estimator can achieve lower \emph{maximum} risk on the support of $\mu$ (since $\rhohat_\mu(D)$'s risk is constant),
\item And therefore no estimator can achieve lower maximum risk over all states (since ``all states'' is a superset of $\mu$'s support).
\end{itemize}
\item Such a stationary measure can occur in one of two ways.  Either $\rhohat_\mu(D)$ has constant risk on \emph{all} states, or $\mu$ is supported on a discrete set.  (Because $\dbar(\rho)$ is analytic, it cannot be constant over a limited range, which means either it is constant everywhere or it has discrete maxima).
\end{enumerate}

\section{Numerical recipes}

\begin{figure}[ht]
\centering
\includegraphics[width=.45\textwidth]{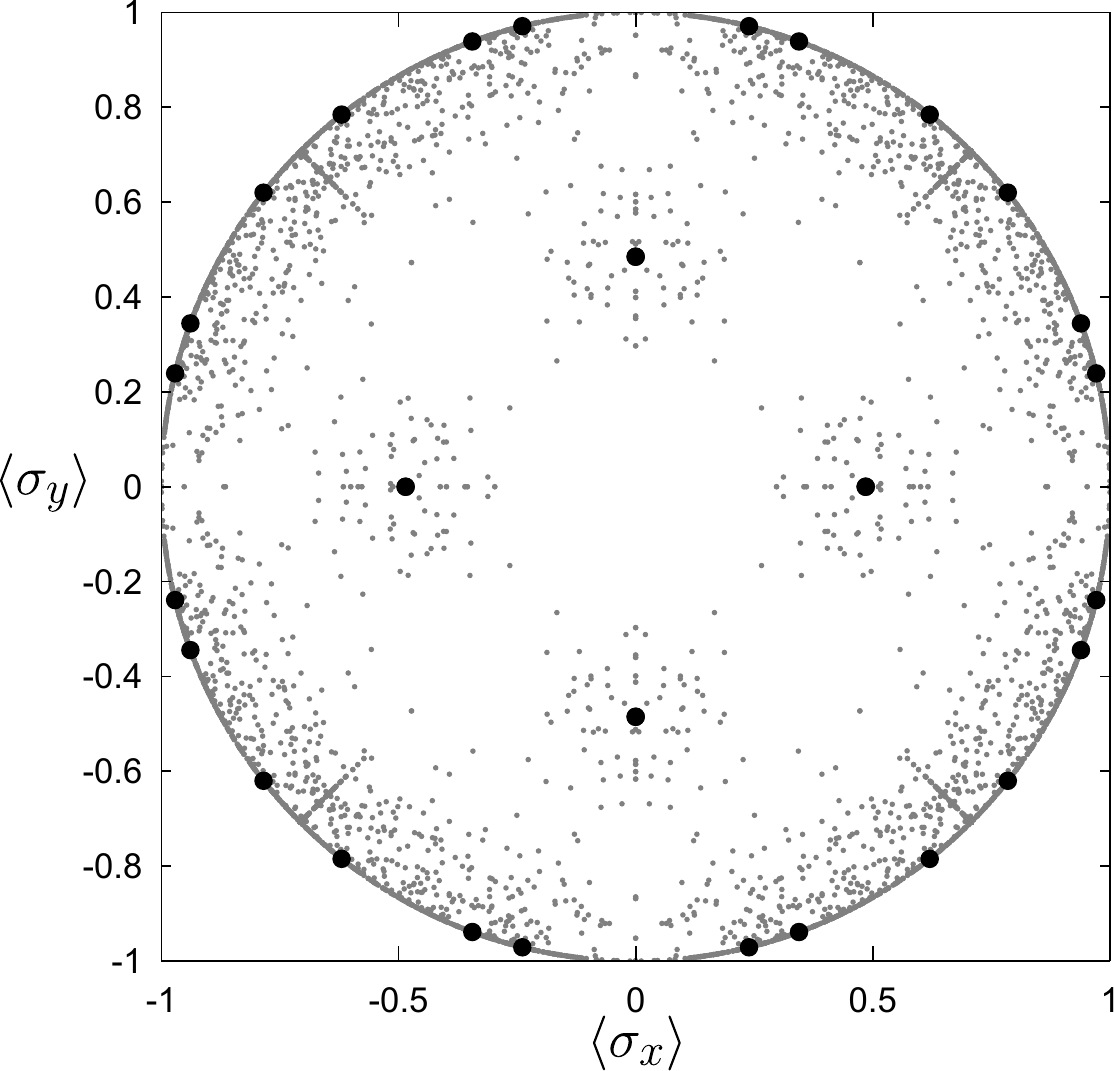}
\includegraphics[width=.45\textwidth]{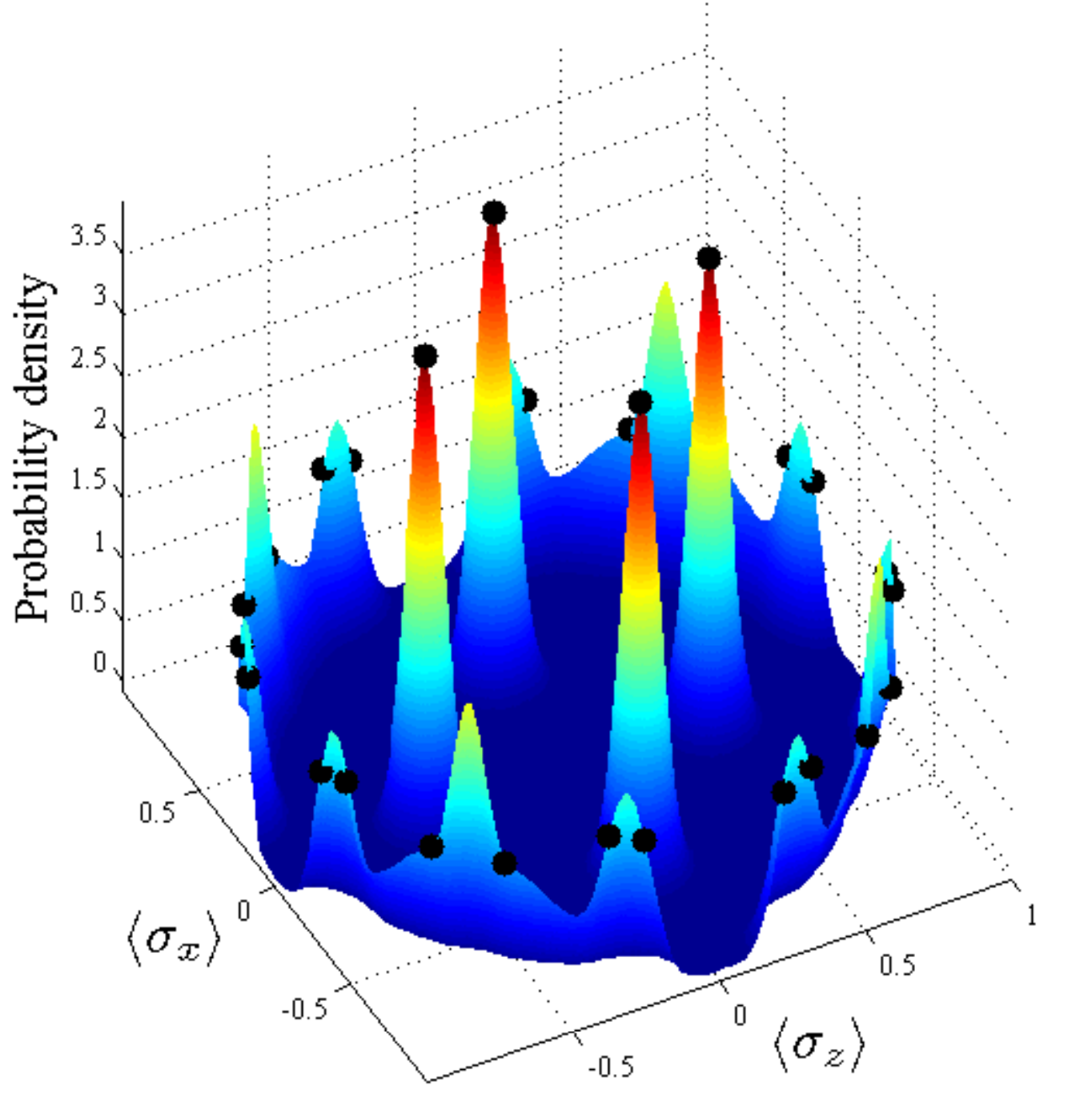}
\caption{\label{fig:2LFPs}
Here we show the support points of numerical approximations to least favorable priors (LFPs) for $N=16$ ($M=8$) Pauli measurements on a \emph{rebit} (a qubit with the constraint $\expect{\sigma_y}=0$.  The weights on these points are not uniform, but we shown a Gaussian kernel density estimate of them on the right.  The LFP found using the highly accurate Algorithm \ref{alg:kempthorne} is supported on the large black dots, while the one found using the much faster Algorithm \ref{alg:mc} is supported on the smaller gray dots.  Note that in this case (and all others where we could use Algorithm \ref{alg:kempthorne}), while the LFPs are evidently different, the resulting minimax risks are indistinguishable.  We conclude that the maximum risk is insensitive to certain visible variations in the prior.}
\end{figure}

The above argument is the basis for the numerical algorithm we used to compute LFPs, and hence the minimax estimators.  Our results were generated using an implementation of Algorithm \ref{alg:kempthorne}, a variant of the one given by Kempthorne \cite{Kempthorne1987Numerical}.  Although this algorithm is drastically more efficient than the brute force approach, it is still insufficient to extract the asymptotic form of the scaling for the risk of minimax estimator.  In particular, it takes $\sim$week to compute the LFP shown in Figure \ref{fig:2LFPs} using an implementation of Algorithm \ref{alg:kempthorne} in MATLAB 2011a on four 2.2GHz processors.  

To remedy this, we devised an efficient Monte Carlo algorithm (Algorithm \ref{alg:mc}).  The core insight is that varying only the weights of the prior renders maximizing the Bayes risk a convex optimization problem.   The algorithm proceeds by randomly choosing $n$ states according to the Hilbert-Schmidt prior.  Then, the Bayes risk is maximized keeping the location of the states fixed.  Both upper and lower bounds on the minimax risk can be obtained.  If these are not close, then we resample near those points whose weights have not be set to zero and repeat the process.

\begin{figure}[ht]
\centering
\includegraphics[width=.45\textwidth]{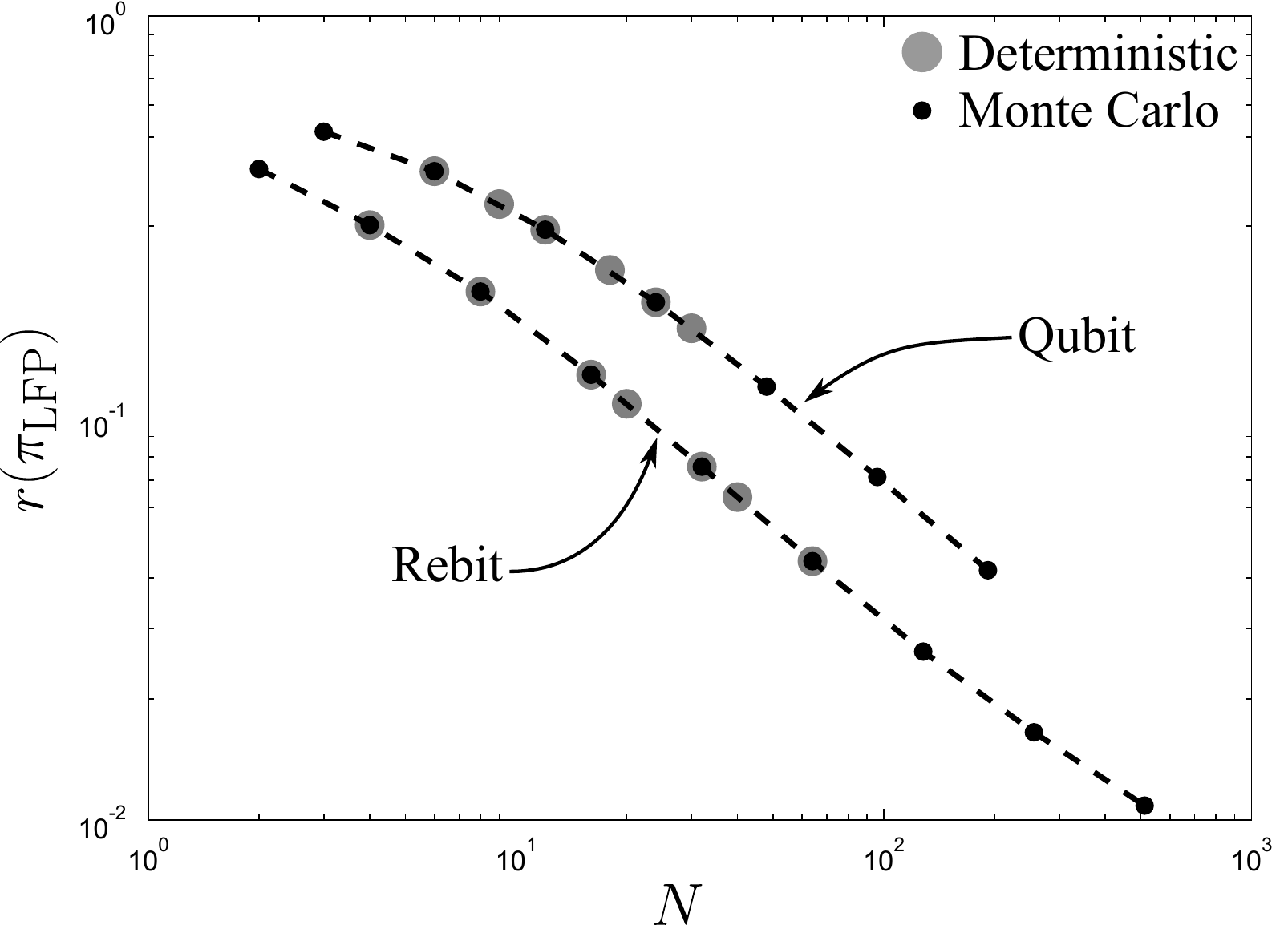}
\caption{\label{fig:compare1}
Comparison of the minimax risk computed using Algorithms \ref{alg:kempthorne} and \ref{alg:mc}.  The minimax risk of qubit and rebit tomography with $N$ Pauli measurements ($N=3\ldots192$ for qubits and $N=2\ldots512$ for rebits) was computed by finding least favorable priors, using Algorithm \ref{alg:kempthorne} (large black dots) and Algorithm \ref{alg:mc} (small gray dots).  In all cases where both algorithms could be applied, results agreed to high precision.}
\end{figure}

Least favorable priors produced by Algorithm \ref{alg:mc} are noticeably different from the (more) exact solutions obtained by Algorithm \ref{alg:kempthorne} (Fig. \ref{fig:2LFPs}).  However, the corresponding Bayes estimators are nearly identical, and these LFPs yield very tight upper and lower bounds on $\dbarmax$ (see Figure \ref{fig:compare1}).  We conclude that the minimax risk is very insensitive to certain variations in the prior.  This explains the discrepancies in the LFPs obtained via Algorithms 1-2, and also justifies our use of the estimators and risks obtained via Algorithm \ref{alg:mc}.  Using this algorithm, we were able to find good approximations to the minimax risk up to $N=192$, but this is still insufficient to clearly show the asymptotic behavior of $\dbarmax$.  For that purpose, we developed the ``noisy coin'' model.
% as shown in Fig.~\ref{fig:NC lower bound}.

\begin{algorithm}
\caption{Kempthorne (deterministic) algorithm for finding the least favorable prior \cite{Kempthorne1987Numerical}.}
\label{alg:kempthorne}
\begin{algorithmic}
\Require Number of measurements $N>0$.
\Require Support points of initial guess prior $x_i$,  $i \in \{1, \dots, n\}$.
\Require Probability weights of the support points $w_i$, $i \in \{1, \dots, n\}$ such that $\sum_i w_i = 1$.
\Require Tolerance $\mathtt{tol}>0$.
\Require Mixing parameter $\alpha$.
\Ensure   Least favorable prior $\{\vec{x},\vec{w}\}$ with $m>n$ support points.
\Ensure   Lower bound on the minimax risk $\mathtt{av\_risk}$.
\Ensure   Upper bound on the minimax risk $\mathtt{max\_risk}$.
\Function{DeterministicLFP}{$N$, $\{\vec{x},\vec{w}\}$,$\mathtt{tol}$}
  \State $\mathtt{diff} \gets \mathtt{tol}+1$
  \While{$\mathtt{diff} > \mathtt{tol}$}
    \State $\{\vec{x},\vec{w}\} \gets$ prior with same number of support points which maximizes the Bayes risk
    \State $\mathtt{av\_risk} \gets$ the maximum value of the Bayes risk for the prior found above
    \State $\mathtt{max\_risk} \gets$ global maximum of risk using the Bayes estimator of the $(\{\vec{x},\vec{w}\})$
    \State $\mathtt{diff}\gets |\mathtt{av\_risk}-\mathtt{max\_risk}|/\mathtt{av\_risk}$
    \If{$\mathtt{diff} > \mathtt{tol}$}
        \State Add a new support where the maximum risk is attained 
        \State $w_{\mathtt{length}(\vec{x})}\gets\alpha$ 
        \State for each $i\leq\mathtt{length}(\vec{x})-1$, $w_i\gets w_i-\alpha/(\mathtt{length}(\vec{x})-1)$   
    \EndIf 
  \EndWhile
  \State \Return $\{\vec{x},\vec{w}\},\mathtt{av\_risk},\mathtt{max\_risk}$
\EndFunction
\end{algorithmic}
\end{algorithm}

\begin{algorithm}
\caption{Monte Carlo algorithm for finding the least favorable prior.}
\label{alg:mc}
\begin{algorithmic}
\Require Number of measurements $N>0$.
\Require Number of support points $n>0$.
\Require Tolerance on accuracy $\mathtt{tol}>0$.
\Require Tolerance on the weights to remove supports $\mathtt{weight\_tol}>0$.
\Require Number of support points to add at each iteration $m>0$ for each current support point.
\Require Variance of normal distribution to sample new points from $\sigma$.
\Ensure   Least favorable prior $\{\vec{x},\vec{w}\}$ with $m>n$ support points.
\Ensure   Lower bound on the minimax risk $\mathtt{av\_risk}$.
\Ensure   Upper bound on the minimax risk $\mathtt{max\_risk}$.
\Function{MCLFP}{$N$, $n$,$\mathtt{tol}$,$\mathtt{weight\_tol}$}
  \State $\mathtt{diff} \gets \mathtt{tol}+1$
  \State $\{\vec{x},\vec{w}\} \gets$ uniform distribution ($w_i=1/n$) sampled according to uniform distribution over $\vec{x}$
  \While{$\mathtt{diff} > \mathtt{tol}$}
    \State $\vec{w} \gets$ weights which maximize the Bayes risk keeping the support points $\vec{x}$ fixed
    \State $\mathtt{av\_risk} \gets$ the maximum value of the Bayes risk for the prior found above
    \State $\mathtt{max\_risk} \gets$ global maximum of risk using the Bayes estimator of the $(\{\vec{x},\vec{w}\})$
    \State $\mathtt{diff}\gets |\mathtt{av\_risk}-\mathtt{max\_risk}|/\mathtt{av\_risk}$
    \State Remove all $x_i$ such that $w_i<\mathtt{weight\_tol}$
    \If{$\mathtt{diff} > \mathtt{tol}$}
      \For{each $x_i$ left}
        \State Add $m$ new support sampled randomly from $\mathcal N (x_i,\sigma)$
      \EndFor
     \State each $w_i\gets 1/\mathtt{length}(\vec{x})$   
    \EndIf 
  \EndWhile
  \State \Return $\{\vec{x},\vec{w}\},\mathtt{av\_risk},\mathtt{max\_risk}$
\EndFunction
\end{algorithmic}
\end{algorithm}


\begin{thebibliography}{16}
\bibitem{foot:errors} Technically, the risk can be written as a sum of terms due to (i) spectral error, and (ii) eigenbasis error.  It turns out that the minimax risk is always dominated by spectral error.  Thus, giving up the eigenbasis error does not lower the risk substantially.
\end{thebibliography}
\end{document}